# UNE CHAINE DE MESURE PERMETTANT DE CARACTERISER LA PERFORMANCE EN VOILE

**Paul Iachkine[1], Sophie Barré[1], Kostia Roncin[2], Jean Michel Kobus[2]**
*1 Service Recherche, Ecole Nationale de Voile, Beg Rohu, 56510 Saint Pierre Quiberon*
*2 LMF UMR Ecole Centrale de Nantes-CNRS 6598, BP 92101 Nantes cedex 3*
iachkine.paul@wanadoo.fr, sophbarre@wanadoo.fr, Jean-Michel.Kobus@ec-nantes.fr,

**Résumé**. Cet article présente les moyens de mesure et les outils d'analyse développés pour la caractérisation de la performance en voile. Ces moyens comprennent un dispositif de mesure de la position et de l'attitude du voilier, un ensemble de capteurs pour la mesure du vent et des outils destinés aux entraîneurs.
**Mots-clé** : voile, mesure, performance, GPS, vent, environnement.

**Introduction**. Une des difficultés de la recherche dans le domaine du sport est de maintenir un équilibre entre les applications de terrain utiles aux sportifs et à leur encadrement et le travail de recherche proprement dit. L'expérience nous a montré que les mesures sur site et l'analyse de la performance pouvaient être le dénominateur commun qui permet de satisfaire tous les partenaires. Il faut pour cela concevoir les mesures et les outils d'analyse pour qu'ils soient facilement mis en œuvre et que les résultats soient utilisables à plusieurs niveaux, à savoir d'une part fournir rapidement des indications directement exploitables par les experts sportifs et d'autre part alimenter des bases de données de qualité, nécessaires à des analyses plus poussées (amélioration de la connaissance, validation de modèles numériques, mise au point de nouvelles procédures d'analyse, etc.).

## 1. Mesure du comportement du voilier

Un dispositif embarqué a été réalisé pour mesurer les grandeurs physiques nécessaires à l'approche objective et quantifiée de la performance d'un bateau et de son équipage dans des conditions réelles de navigation. Ces mesures sont principalement (figure 1) la trajectoire et la vitesse obtenues grâce à un GPS différentiel submétrique, le cap, la gîte et l'assiette et toutes les données dynamiques obtenues par une centrale inertielle. Le système permet actuellement la mesure de l'angle de barre mais pourra enregistrer d'autres grandeurs caractérisant le fonctionnement et la conduite du bateau.

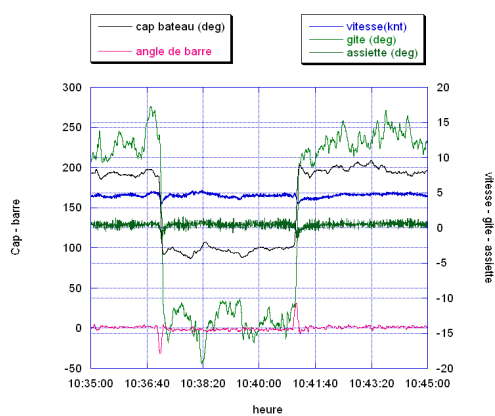

*Figure 1 : exemple d'enregistrements de données des différents capteurs.*

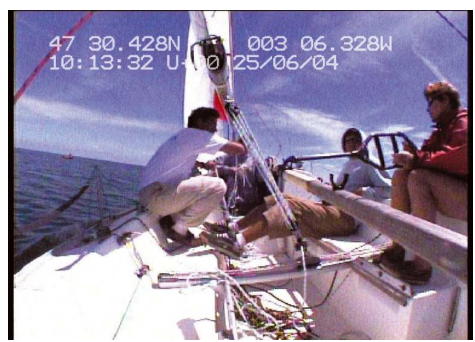

*Figure 2 : vue de la caméra embarquée.*

L'architecture du système est basée sur une carte PC104, un bus CAN et un modem UHF. Ce système permet de gérer les capteurs, de stocker les données brutes et de transmettre une partie de ces données pour un contrôle à distance. Une station DGPS à terre permet un post-traitement des données GPS pour obtenir une précision inférieure au décimètre. Une caméra embarquée et des microphones complètent le dispositif (figure 2). Ces données audio-vidéo fournissent des informations sur la position de l'équipage dans le bateau et les réglages utilisés. L'incrustation de l'heure GPS sur les images vidéo permet de les synchroniser avec les autres mesures.

## 2. Mesure de l'environnement

Les grands organismes de la météorologie et de l'océanographie réalisent des mesures du vent, du courant et de l'état de la mer soit régulièrement soit lors de campagnes spécifiques. Cette activité se caractérise par l'ampleur des moyens mis en œuvre et par le fait qu'elles s'intéressent à des échelles de temps et d'espace beaucoup plus grandes que celles qui sont utiles pour l'étude des régates côtières olympiques ou de match racing. Même si les modèles et les données fournies aux échelles régionales et locales sont d'un intérêt évident pour la pratique de la voile, il est nécessaire de les compléter par des moyens et des méthodes d'analyse adaptées aux micro-échelles de quelques centaines de mètres.

Pour ce faire nous utilisons actuellement quatre systèmes mobiles (petits catamarans de sport) qui permettent des mesures de vent en quatre points du plan d'eau étudié, avec une transmission par liaison UHF des données vers un ordinateur à terre. Chaque système est composé d'un anémomètre à ultrason, d'un compas électronique, d'un GPS et d'un modem UHF. Ces mesures sont complétées par un point de mesure fixe (vent et pression atmosphérique) directement relié à cet ordinateur.

A partir de ces quatre points de mesure, le vent est interpolé pour chaque instant et chaque point du plan d'eau. Cette méthode permet de connaître le vent que reçoit le bateau sans l'instrumenter et donc de fournir une mesure normalisée du vent quels que soient la taille et le type de bateau. La méthode a également l'avantage de palier les inconvénients d'une mesure du vent relatif sur le bateau sensiblement perturbée par l'écoulement autour des voiles et par les mouvements [1].

Enfin, les données enregistrées sur les plans d'eau seront utilisées pour caractériser l'évolution spatio-temporelle du vent et pour valider les modèles micro-météorologiques en cours d'élaboration.

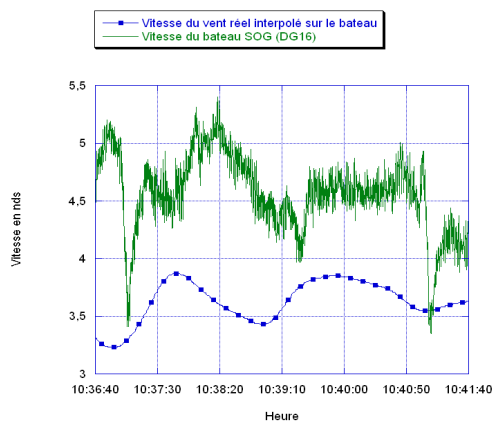

*Figure 3 : évolutions comparées de la vitesse du bateau mesurée et de la force du vent obtenue par interpolation.*

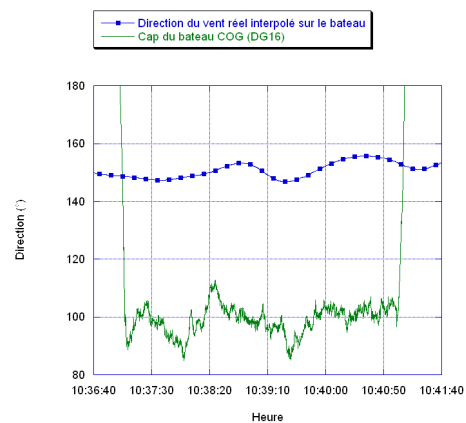

*Figure 4 : évolutions comparées du cap du bateau mesuré et de la direction du vent obtenue par interpolation.*

Les figures 3 et 4 montrent la bonne corrélation entre ce vent interpolé et les performances du bateau.

Les valeurs de courant ne sont pas, pour l'instant, mesurées mais obtenues à partir de la base de données du SHOM. De même, l'état de la mer n'est pas mesuré mais apprécié « visuellement ». Pour ces deux grandeurs, un outil adapté à nos besoins est en cours d'étude au laboratoire de Mécanique des Fluides de l'ECN.

Le logiciel MAXSEA permet à l'entraîneur de visualiser l'ensemble de ces données environnementales par-dessus le fond de carte et en ajoutant le relief si nécessaire ou les trajectoire des bateaux (figure 5).

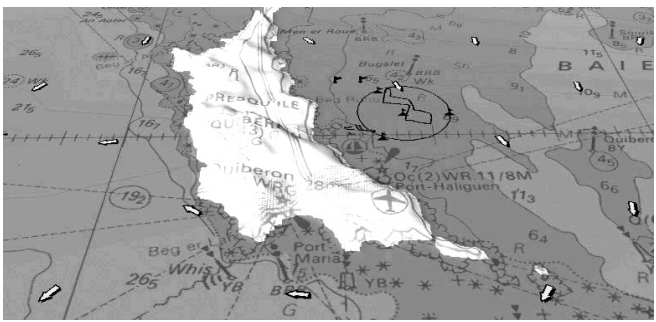 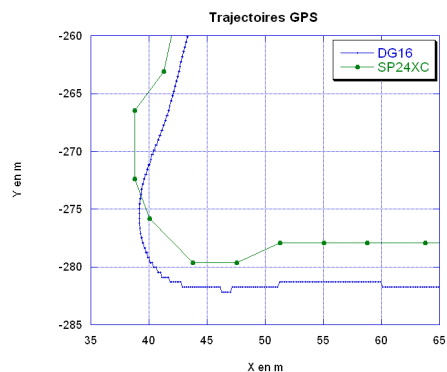

*Figure 5 : exemple de visualisation.*  *Figure 6 : Comparaison des données entre DGPS professionnel (20Hz) et GPS grand public (0,5Hz).*

### 3. Les outils pour l'entraîneur

Depuis 2002, les GPS sont utilisés couramment et de façon autonome comme outil d'aide à l'entraînement au même titre que la caméra vidéo. L'opportunité du développement d'un modèle portable, réalisé pour la Fédération de Vol Libre et disposant d'une capacité de mémoire étendue, a permis cette utilisation. La précision de ces GPS "grand public" (figure 6) est suffisante pour une analyse tactique même si elle est encore insuffisante pas une analyse fine des manœuvres ou de la conduite (technique de barre). Sur chaque bateau, un GPS portable est installé et récupéré à la fin de la navigation par l'entraîneur.

Le logiciel NAVIRACE, développé à l'initiative de l'Ecole Nationale de Voile, permet de transférer les données des traces sur un PC, d'analyser les trajectoires pendant les débriefings (fonctions magnétoscope) et de les exporter ensuite vers un logiciel de cartographie (Maxsea).

Un second outil, baptisé VISUPERF, est en développement, il s'agit d'un logiciel permettant une visualisation synchronisée de la vidéo et des mesures. Un démonstrateur sous MATLAB a été réalisé dans le cadre d'un DESS, il doit être validé courant 2005 par plusieurs entraîneurs.

### 4. Collaborations

Ces outils sont développés dans le cadre d'un projet fédérant l'Université de Nantes (ECN, IUT de Nantes et Ecole Polytechnique de Nantes), un Etablissement National du Ministère des Sports (ENV) et des industriels (Thales Navigation, Cadden, LCJ capteurs et Maxsea). Ils seront utilisés par l'Equipe de France pour l'étude des performances de la nouvelle planche olympique.

### 5. Perspectives

Une campagne d'essais mettant en œuvre le matériel de mesure du comportement du voilier et du vent a été réalisée en juin 2004. Elle a permis d'évaluer les performances et les fonctionnalités. Les résultats sont actuellement utilisés pour valider les simulations numériques du fonctionnement du bateau obtenues par le simulateur développé par Kostia Roncin [2]. L'objectif est maintenant de fiabiliser le matériel, et de réaliser deux systèmes complets pour le match racing. Les travaux sur la micro-météorologie permettront d'améliorer l'interpolation du vent sur les plans d'eau. Pour ce qui concerne les outils mis à disposition des entraîneurs, l'objectif est d'améliorer la précision du positionnement à la fois en exploitation directe et en post-traitement et de poursuivre le développement des logiciels d'exploitation de ces mesures.

### Références